\documentclass[preprint]{aastex}

\begin{document}

\title {The Pulsating Eclipsing Binary TIC 309658221 in a 7.59 Day Orbit }
\author{Jae Woo Lee$^{1}$, Martti H. Kristiansen$^{2,3}$, and Kyeongsoo Hong$^{4}$  }
\affil{$^1$Korea Astronomy and Space Science Institute, Daejeon 34055, Korea}
\affil{$^2$DTU Space, National Space Institute, Technical University of Denmark, Elektrovej 327, DK-2800 Lyngby, Denmark}
\affil{$^3$Brorfelde Observatory, Observator Gyldenkernes Vej 7, DK-4340 T\o{}ll\o{}se, Denmark}
\affil{$^4$Institute for Astrophysics, Chungbuk National University, Cheongju 28644, Korea}
\email{jwlee@kasi.re.kr}

\begin{abstract}
We present a new eclipsing binary (EB) showing multiperiodic oscillations using the first three sectors of {\it TESS} photometry. 
The eclipse and pulsation light curves of TIC 309658221 were modeled using an iterative method to obtain a consistent photometric solution. 
The {\it TESS} target is a circular-orbit, detached binary system with a mass ratio of 0.349, an inclination angle of 80.42 deg, 
and a temperature difference of 847 K between the components. The primary component of the system lies near the red edge of 
the $\delta$ Sct instability region on the main-sequence band in the Hertzsprung-Russell diagram. Multiple frequency analyses were 
applied to the eclipse-subtracted residuals after removing the binary effects in the observed data. These resulted in the detection 
of 26 frequencies, of which $f_1-f_6$ were independent pulsation frequencies. The 20 other frequencies could be mainly caused by 
orbital harmonics ($f_8$ and $f_{11}$) or combination frequencies. The period ratios and pulsation constants of the $f_1-f_6$ frequencies 
are in the ranges of $P_{\rm pul}/P_{\rm orb}$ = 0.010$-$0.013 and Q = 0.027$-$0.036 days, respectively, which are typical of $\delta$ Sct type. 
The results reveal that TIC 309658221 is an eclipsing $\delta$ Sct star with an orbital period of 7.5952 days and pulsation frequencies 
of 9.94$-$13.01 day$^{-1}$. This work demonstrates that the 2-min cadence observations of {\it TESS} are very useful for the study of 
pulsating EBs with multiple frequencies and low amplitudes.
\end{abstract}

{}

\section{INTRODUCTION}

Near-continuous and ultra-precise photometric data have been obtained from space missions, such as {\it CoRot} (Baglin et al. 2006; 
Auvergne et al. 2009) and {\it Kepler} (Borucki et al. 2010; Koch et al. 2010). Although designed to search for extrasolar planets, 
these systems have revolutionized variable star studies, especially asteroseismology. The time-series data from these missions allow 
the detection of many pulsation frequencies with amplitudes lower than 0.1 mmag. Launched on 2018 April 18, the Transiting Exoplanet 
Survey Satellite ({\it TESS}) is a full-sky photometric survey to find small planets orbiting nearby stars 10$-$100 times brighter 
than those observed by {\it Kepler} (Ricker et al. 2015). It will survey over 85 \% of the sky in a highly elliptical 13.7-day orbit 
during the 2-year primary mission. In this survey, each ecliptic hemisphere is divided into 13 partially overlapping sectors of 
24 deg $\times$ 96 deg, each of which is continuously observed for two {\it TESS} orbits of 27.4 days. The satellite covers 
the southern ecliptic hemisphere in the first year and then observes the northern hemisphere in the following year. 

EBs allow an accurate and direct determination of the masses and radii of stars, and pulsating stars allow stellar interiors to be probed 
from core to surface layers using asteroseismology. Then, the strong synergy between EBs and pulsators can play a critical reciprocal role 
in providing significant information on the stellar parameters and internal structures. We have been searching for and characterizing 
the pulsating EBs using {\it Kepler} photometry (Lee et al. 2014, 2019) and ground-based spectroscopy (Hong et al. 2015; Lee \& Park 2018). 
In this paper, our program target is TIC 309658221 (HD 34954, TYC 8878-116-1, GSC 8878-0116, 2MASS J05170089-6130524; 
$V\rm_T$ = $+$10.31, $(B-V)\rm_T$ = $+$0.35; H\o g et al. 2000), which was known as a single F-type star prior to the {\it TESS} 
observations (Stassun et al. 2018). Houk \& Cowley (1975) classified the spectral type of the star as F2 III/IV but the classification 
might be deduced from the combined spectrum of the binary components. We show that this object is a new eclipsing binary with 
multiperiodic pulsations in a 7.59-day orbit.

\section{{\it TESS} PHOTOMETRY AND ORBITAL PERIOD}

The {\it TESS} with 2-s exposures provides 2-min cadence observations of more than 200,000 preselected stars and full-frame images of 
24 deg $\times$ 96 deg recorded every 30 min. TIC 309658221 was observed in a 2-min cadence mode from 2018 July 25. We used 
the simple aperture photometry data obtained during Sectors 1, 2, and 3 in camera 4 available at MAST\footnote{https://archive.stsci.edu/}. 
During our analysis, additional data for TIC 309658221 were released for Sections 4, 5 and 6 and we further note a continuous data stream 
in the southern hemisphere (Sector 1-13) using the Web {\it TESS} Viewing Tool\footnote{https://heasarc.gsfc.nasa.gov/cgi-bin/tess/webtess/wtv.py}.
The uncorrected flux measurements were detrended and normalized following a procedure similar to that described by Lee et al. (2017). 
These were converted to a magnitude scale by applying a {\it TESS} magnitude of $+$9.84 (Stassun et al. 2018) at maximum light. 
We removed outliers that displayed quite a big deviation compared to neighboring observations by visual inspection, instead of 
automated sigma clipping because of the presence of the oscillation signal. As an example, the resultant light curve for Sector 1 is 
shown in Figure 1, where both eclipses and pulsations are clearly visible. 

In order to obtain an orbital ephemeris for our program target, the times of minimum light and their errors were measured from the {\it TESS} 
observations by using the method of Kwee \& van Woerden (1956). We applied a linear least-squares fit to the minimum times and determined 
the following ephemeris:
\begin{equation}
 C_1 = \mbox{BJD}~ 2,458,365.204170(73) + 7.594648(23)E.
\end{equation}
The parenthesized quantities are the 1$\sigma$ uncertainty for the last digit of each term. Individual timing errors were used as 
weights. These are given in columns (1)$-$(5) of Table 1, where we present the $O$--$C_1$ values and epochs computed using 
equation (1). As shown in Figure 1 and discussed in the following sections, because TIC 30968221 is a pulsating EB exhibiting 
mutiperiodic oscillations, the eclipse times may be affected by the light variations. Hence, we recalculated the minimum times 
from each eclipse curve after subtracting all frequencies detected in Section 4 from the original {\it TESS} data. A linear fit to 
the new timings resulted in an improved ephemeris, as follows:
\begin{equation}
 C_2 = \mbox{BJD}~ 2,458,365.206340(52) + 7.595165(17)E.
 \end{equation}
The results are presented in columns (6)$-$(8) of Table 1. The timing residuals of the ephemeris (2) show a standard deviation 
of $\pm$0.0024 d, which is over two times smaller than that ($\pm$0.0064 d) of equation (1).

\section{BINARY MODELING}

A total of 52,776 individual observations from Sector 1 (BJD 2458325.29$-$2458353.17), Sector 2 (BJD 2458354.10$-$2458381.52), 
and Sector 3 (BJD 2458382.03$-$2458409.38) were modeled with the Wilson-Devinney synthesis code (Wilson \& Devinney 1971, 
van Hamme \& Wilson 2007; hereafter W-D). The effective temperature of the hotter primary star was set to be 6993 $\pm$ 200 K from 
Gaia DR2\footnote{https://gea.esac.esa.int/archive/} (Gaia Collaboration et al. 2018). We used the logarithmic limb-darkening coefficients 
taken from the tables of van Hamme (1993). Until now TIC 309658221 was known as a single star and no binary parameters exist. Thus, 
the binary modeling of the system was made in a way almost identical to that for the pulsating EBs KIC 11401845 (Lee et al. 2017) and 
EPIC 245932119 (Lee et al. 2019), applying the mass ratio ($q$)-search procedure to modes 2, 3, 4 and 5 of the W-D code. 
The differential correction (DC) program of the W-D code was carried out on each assumed mass ratio and the four modes, but only 
the light curve synthesis for detached mode 2 seemed to be acceptable for TIC 309658221, which had an optimal solution at $q$ = 0.35. 

From that time on, this value of $q$ was treated as an adjustable parameter in deriving the light curve parameters of this binary. 
The DC program was run until the corrections to the parameters became smaller than the corresponding standard deviations. The result 
is listed in Model 1 of Table 2. In order to estimate more reliable errors for the fitted parameters, the observed {\it TESS} data were 
divided into ten subsets and analyzed individually with the W-D code. Then, we computed the 1$\sigma$-values of each parameter from 
the ten different values (Koo et al. 2014). These values are given as parenthesized errors in Table 2. The synthetic light curve from 
Model 1 is given as a blue solid curve in the top panel of Figure 2 and the corresponding residuals are plotted in the middle panel of 
this figure. In all procedures including the $q$ searches, we adjusted an orbital eccentricity ($e$) and a third light ($l_3$) as 
additional parameters, but their values remained indistinguishable from zero. 

Assuming that the primary component is a normal main-sequence star, its surface temperature corresponds to a spectral type of F1$\pm$1 
and a mass of $M_1$ = 1.50$\pm$0.08 $M_\odot$, based on their mutual relationship\footnote{http://www.pas.rochester.edu/$\sim$emamajek/EEM\_dwarf\_UBVIJHK\_colors\_Teff.txt} 
(Pecaut \& Mamajek 2013). We computed the absolute dimensions for TIC 309658221 from the $M_1$ value and the light curve parameters 
in Table 2. These are presented in the lower part of Table 2. We calculated the absolute visual magnitudes ($M_{\rm V}$) by using 
the bolometric corrections (BCs) of Torres (2010) between $\log T_{\rm eff}$ and BC. The first distance to the EB system was determined 
to be 482 pc from an apparent magnitude of $V$ = +10.276 (H\o g et al. 2000) and the interstellar reddening of $A_{\rm V}$ = 0.078 
(Schlafly \& Finkbeiner 2011).

\section{PULSATIONAL CHARACTERISTICS}

In Figure 3, the light curve residuals were plotted as the magnitudes {\it versus} BJDs. We could see oscillations with a semi-amplitude 
of about 6 mmag maximum and varying from cycle to cycle. To detect multiple frequencies in an iterative method (Lee et al. 2017, 2019), 
we applied frequency analyses to these residuals. The PERIOD04 program (Lenz \& Breger 2005) was performed up to the Nyquist limit 
of $f_{\rm Ny}$ = 360 day$^{-1}$. With the successive prewhitening process (Lee et al. 2014), we found the frequencies with the signal 
to noise amplitude ratio (S/N) larger than the empirical criterion of 4.0 proposed by Breger et al. (1993). Then, we analyzed 
the pulsation-subtracted data with the W-D code after removing the pulsation signatures from the original {\it TESS} data. New light curve 
parameters were used to obtain the corresponding eclipse-subtracted residuals, which were again introduced into the PERIOD04 program for 
multiple frequency analyses. 

The above procedure was repeated until no significant signal was detected. The final light curve parameters and absolute dimensions are 
listed in Model 2 of Table 2, which are in good agreement with those from Model 1 within their errors. This indicates that the pulsations 
do not have much effect on the photometric solution of TIC 309658221. The pulsation-subtracted data and synthetic light curve are plotted 
as black circles and a red line, respectively, in the top panels of Figure 2. From the detailed analysis, we detected 26 frequencies with 
S/N $>$ 4.0 listed in Table 3. The uncertainties for the frequency determination were calculated following the treatment of 
Kallinger et al. (2008). The result from PERIOD04 is shown in Figure 4. The amplitude spectra after prewhitening 
the first six frequencies and then all 26 frequencies are presented in the second and bottom panels of Figure 4, respectively. 
The synthetic curve computed from the 26-frequency fit appears as a red line in the lower panel of Figure 3. We plotted the combination 
of the two effects (eclipses and pulsations) as green solid curves in Figure 1. 

As shown in Figure 4 and Table 3, the main frequency peaks of TIC 309658221 lie in a region between 9 and 14 day$^{-1}$ and 
additional signals are visible at very low frequencies of $< 1$ day$^{-1}$. We examined the frequencies to identify possible harmonic 
and combination terms within the resolution of 1.5/$\Delta t$ = 0.019 day$^{-1}$, where $\Delta t$ = 81 days is the time base of 
observations (Loumos \& Deeming 1978). The result is presented in the last column of Table 3. The first six frequencies ($f_1-f_6$) are 
pulsation frequencies, and the $f_8$ and $f_{11}$ frequencies appear to be the orbital frequency of $f_{\rm orb}$ = 0.13166 day$^{-1}$ 
and its three times, respectively. The two low frequencies could be partially attributed to the imperfect removal of the binary effects 
in the {\it TESS} data. The majority of the remaining 18 frequencies mainly arise from combination frequencies.

\section{SUMMARY AND DISCUSSION}

TIC 309658221 was observed during Sectors 1, 2, and 3 of the {\it TESS} operations. The 2-min cadence observations exhibited 
both eclipses and oscillations, which implies that this object is not a single star but a pulsating EB. To improve binary and 
pulsation solutions, we analyzed the observed and prewhitened (eclipse- and pulsation-subtracted) data in an iterative process. 
The binary modeling indicates that TIC 309658221 is in a circular-orbit, detached configuration, where the primary and 
secondary components fill 27 \% and 47 \% of their inner Roche lobe, respectively. We estimated the absolute dimensions of 
each component from the photometric solutions and the empirical relations between effective temperature (spectral type) and 
stellar mass. As in previous studies (Lee et al. 2017; Lee \& Park 2018), the binary parameters are immune to the oscillations. 
In mass-radius, mass-luminosity, and Hertzsprung-Russell (HR) diagrams (Lee et al. 2016; Lee \& Park 2018), the primary component 
is a main-sequence star located near the red edge of the $\delta$ Sct instability strip. The low-mass secondary is highly 
oversized and overluminous, compared with dwarf stars of the same mass, but is not unusual for companions in short binaries 
(e.g., Table 1 of Liakos \& Niarchos 2017). 

We removed the binary effects in the observed {\it TESS} data and applied multiple frequency analyses to the light curve residuals. 
Twenty-six frequencies with S/N $>$ 4.0 were detected in the two ranges of $< 1$ day$^{-1}$ and 8.0$-$20.5 day$^{-1}$. As shown in 
Table 3, $f_1-f_6$ are independent pulsation frequencies, while the other frequencies may be orbital harmonics and combination terms. 
Using the Model 2 parameters in Table 2 and the equation of 
$\log Q_i = -\log f_i + 0.5 \log g + 0.1M_{\rm bol} + \log T_{\rm eff} - 6.456$ (Petersen \& J\o rgensen 1972), we computed 
the observed pulsation constants of the six frequencies to be  $Q_1$ = 0.036 days, $Q_2$ = 0.034 days, $Q_3$ = 0.033 days, 
$Q_4$ = 0.027 days, $Q_5$ = 0.029 days, and $Q_6$ = 0.034 days. The pulsation periods ($P_{\rm pul}$) of 0.077$-$0.101 days and 
the pulsation constants of 0.027$-$0.036 days correspond to pressure modes of $\delta$ Sct stars (Breger 2000). Further, 
$P_{\rm pul}/P_{\rm orb}$ = 0.010$-$0.013 are within the upper limit of 0.09 for $\delta$ Sct pulsators in binaries (Zhang et al. 2013). 
These values and the position on the HR diagram reveal that the primary component of TIC 309658221 would be a $\delta$ Sct pulsator. 

Our study demonstrates the potential of the 2-min cadence {\it TESS} data to characterize the pulsating stars in EBs. Despite 
our success in modeling TIC 309658221, the binary parameters are preliminary because of the absence of spectroscopic data. 
The pulsating EB is a relatively bright star with a long orbital period, so it is possible to carry out spectroscopic follow-up 
observations with 2-m class telescopes and to measure double-lined radial velocities. When high-resolution spectroscopy is undertaken, 
the accurate absolute masses, radii, temperatures, and luminosities can be directly derived and, hence, the physical properties of 
the system will be understood better than is possible with photometry alone.

\acknowledgments{ }
This paper includes data collected by the {\it TESS} mission, which are publicly available from MAST. Funding for the {\it TESS} mission 
is provided by the NASA Science Mission directorate. We acknowledge the use of public {\it TESS} Alert data from pipelines at the TESS 
Science Office and at the TESS Science Processing Operations Center. This research has made use of the Simbad database maintained at CDS, 
Strasbourg, France, and was supported by the KASI grant 2019-1-830-03. K.H. was supported by the grant numbers 2017R1A4A1015178 of 
the National Research Foundation (NRF) of Korea.

\clearpage
\begin{figure}
\includegraphics[scale=0.95]{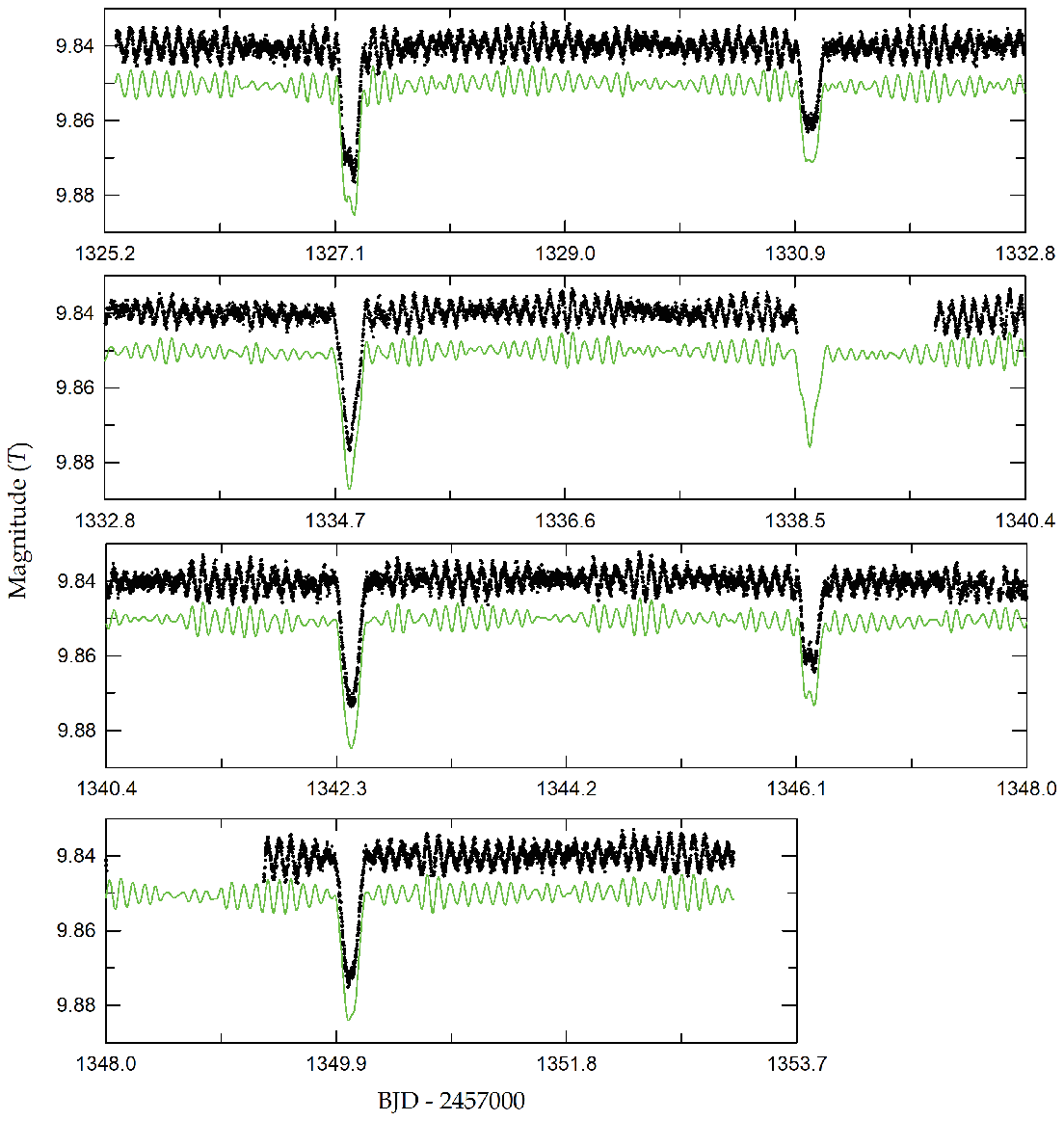}
\caption{Black circles present time-series data of TIC 309658221 observed during the {\it TESS} Sector 1 and separated at intervals of 
7.6 days. The green lines are the sum of the two model curves computed from the Model 2 parameters of Table 2 and the 26-frequency fit 
of Table 3, respectively. They are offset by +0.01 mag for clarity. }
\label{Fig1}
\end{figure}

\begin{figure}
\includegraphics[]{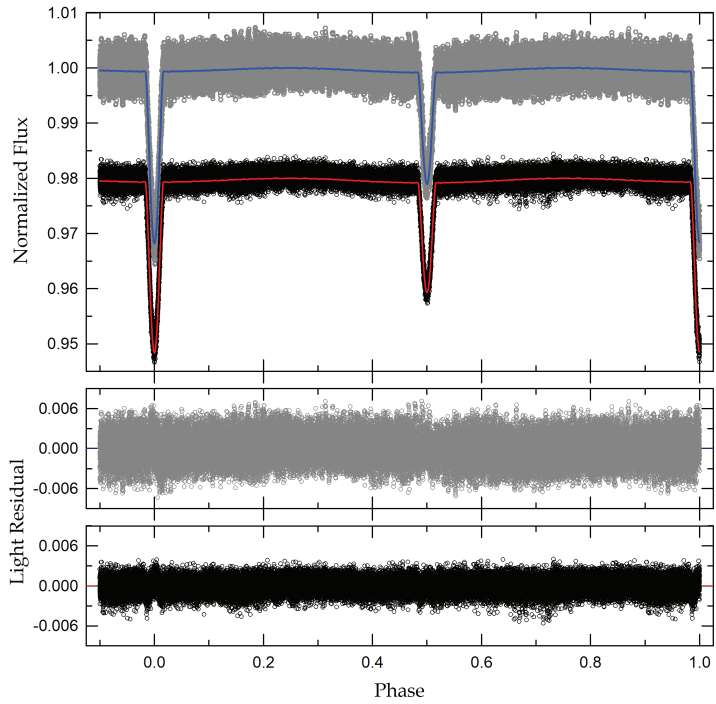}
\caption{Binary light curve before (gray circle) and after (black circle) subtracting the pulsation signatures from the observed {\it TESS} data. 
The blue and red lines are computed with the Model 1 and Model 2 parameters of Table 1, respectively. The corresponding residuals from the fits 
are plotted in the middle and bottom panels in the same order as the light curves. }
\label{Fig2}
\end{figure}

\begin{figure}
\includegraphics[]{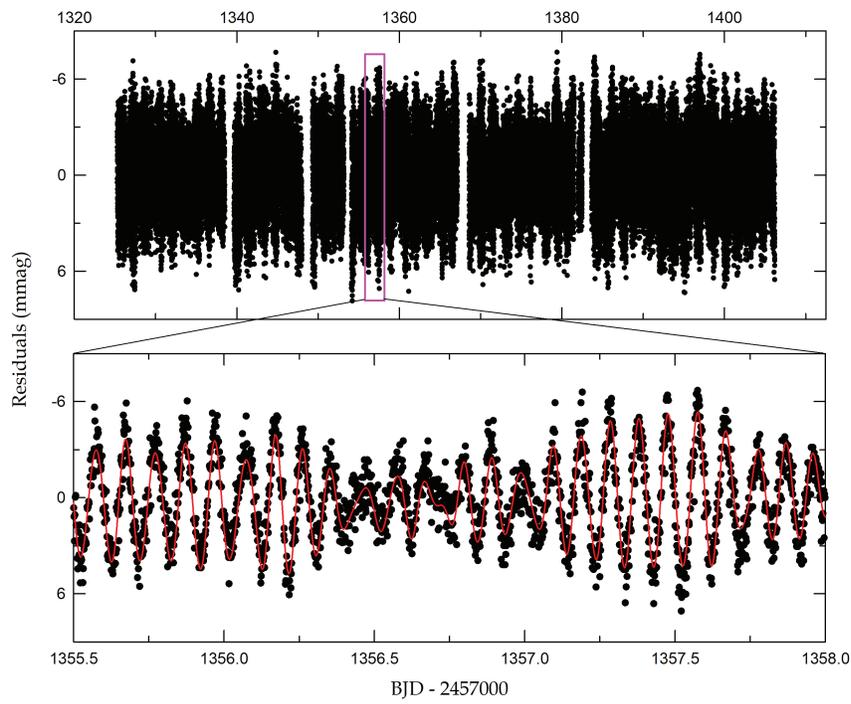}
\caption{Light curve residuals after removing the binarity effect from the observed {\it TESS} data. The lower panel presents a short section 
of the residuals marked by the inset box in the upper panel. The synthetic curve is computed from the 26-frequency fit to the entire residuals. }
\label{Fig3}
\end{figure}

\begin{figure}
\includegraphics[]{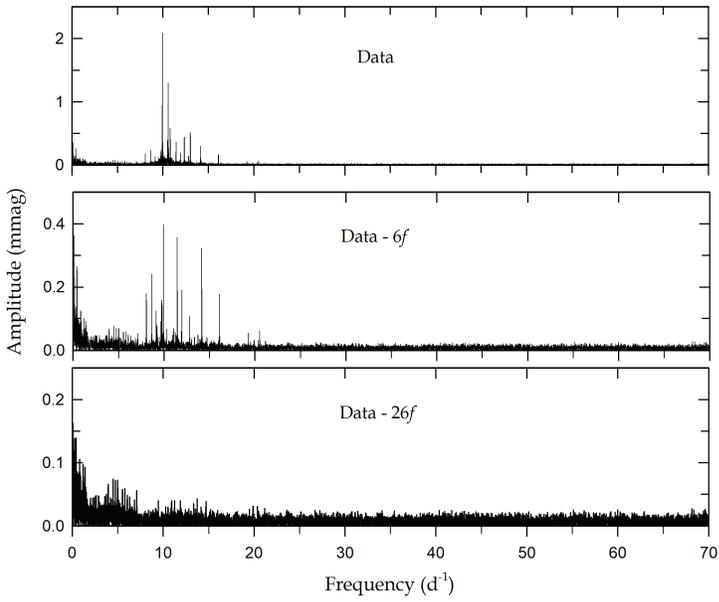}
\caption{Amplitude spectra before (top panel) and after pre-whitening the first six frequencies (second) and all 26 frequencies (bottom) 
from the PERIOD04 program for all light residuals. }
\label{Fig4}
\end{figure}

\clearpage
\begin{deluxetable}{lcccrcclcc}
\tabletypesize{\small}  
\tablewidth{0pt}
\tablecaption{Eclipse Timings Measured from Both Datasets: Including and Removing the Pulsation Signatures }
\tablehead{
\multicolumn{3}{c}{Including Pulsations}                  &&                 &                && \multicolumn{3}{c}{Removing Pulsations}                     \\ [1.0mm] \cline{1-3} \cline{8-10} \\[-2.0ex]
\colhead{BJD}    & \colhead{Error} & \colhead{$O$--$C_1$} && \colhead{Epoch} & \colhead{Min}  && \colhead{BJD}     & \colhead{Error} & \colhead{$O$--$C_2$}   
}                                                                         
\startdata                                                                
2,458,327.22805	 & $\pm$0.00048    & $-$0.00289           && $-$5.0          & I              && 2,458,327.23049	 & $\pm$0.00028    & $-$0.00003	           \\
2,458,331.02846	 & $\pm$0.00016    & $+$0.00021           && $-$4.5          & II             && 2,458,331.02831	 & $\pm$0.00016    & $-$0.00021	           \\
2,458,334.82567	 & $\pm$0.00054    & $+$0.00009           && $-$4.0          & I              && 2,458,334.82586	 & $\pm$0.00026    & $-$0.00018	           \\
2,458,342.42052	 & $\pm$0.00024    & $+$0.00030           && $-$3.0          & I              && 2,458,342.42080	 & $\pm$0.00021    & $-$0.00004	           \\
2,458,346.21746	 & $\pm$0.00074    & $-$0.00009           && $-$2.5          & II             && 2,458,346.22076	 & $\pm$0.00041    & $-$0.00233	           \\	     
2,458,350.01585	 & $\pm$0.00023    & $+$0.00097           && $-$2.0          & I              && 2,458,350.01571	 & $\pm$0.00020    & $-$0.00030	           \\
2,458,357.61243	 & $\pm$0.00137    & $+$0.00291           && $-$1.0          & I              && 2,458,357.61032	 & $\pm$0.00027    & $-$0.00086	           \\
2,458,361.40918	 & $\pm$0.00023    & $+$0.00234           && $-$0.5          & II             && 2,458,361.40929	 & $\pm$0.00022    & $-$0.00053	           \\	       
2,458,365.20931	 & $\pm$0.00087    & $+$0.00514           && $+$0.0          & I              && 2,458,365.20614	 & $\pm$0.00028    & $-$0.00020	           \\
2,458,369.00414	 & $\pm$0.00052    & $+$0.00265           && $+$0.5          & II             && 2,458,369.00416	 & $\pm$0.00021    & $-$0.00023	           \\	         
2,458,372.80036	 & $\pm$0.00063    & $+$0.00154           && $+$1.0          & I              && 2,458,372.80056	 & $\pm$0.00016    & $-$0.00095	           \\
2,458,376.59548	 & $\pm$0.00053    & $-$0.00066           && $+$1.5          & II             && 2,458,376.59942	 & $\pm$0.00023    & $-$0.00033	           \\	       
2,458,380.39662	 & $\pm$0.00027    & $+$0.00315           && $+$2.0          & I              && 2,458,380.39645	 & $\pm$0.00028    & $-$0.00022	           \\
2,458,384.18102	 & $\pm$0.00018    & $-$0.00977           && $+$2.5          & II             && 2,458,384.19322	 & $\pm$0.00043    & $-$0.00103	           \\	       
2,458,387.99114	 & $\pm$0.00043    & $+$0.00302           && $+$3.0          & I              && 2,458,387.99117	 & $\pm$0.00018    & $-$0.00067	           \\
2,458,391.78882	 & $\pm$0.00034    & $+$0.00338           && $+$3.5          & II             && 2,458,391.78684	 & $\pm$0.00024    & $-$0.00258	           \\	         
2,458,395.60799	 & $\pm$0.00059    & $+$0.02523           && $+$4.0          & I              && 2,458,395.59629	 & $\pm$0.00026    & $-$0.00929	           \\
2,458,399.38446	 & $\pm$0.00039    & $+$0.00437           && $+$4.5          & II             && 2,458,399.38237	 & $\pm$0.00027    & $-$0.00221	           \\	         
2,458,403.18110	 & $\pm$0.00036    & $+$0.00369           && $+$5.0          & I              && 2,458,403.18140	 & $\pm$0.00021    & $-$0.00076	           \\
\enddata
\end{deluxetable}

\begin{deluxetable}{lcccccccc}
\tablewidth{0pt} 
\tablecaption{Binary Parameters of TIC 309658221 }
\tablehead{
\colhead{Parameter}                      & \multicolumn{2}{c}{Model 1$\rm ^a$}         && \multicolumn{2}{c}{Model 2$\rm ^b$}         \\ [1.0mm] \cline{2-3} \cline{5-6} \\[-2.0ex]
                                         & \colhead{Primary} & \colhead{Secondary}     && \colhead{Primary} & \colhead{Secondary}                                                  
}                                                                                                                                     
\startdata                                                                                                                            
$q$                                      & \multicolumn{2}{c}{0.3494(43)}              && \multicolumn{2}{c}{0.3491(17)}              \\
$i$ (deg)                                & \multicolumn{2}{c}{80.463(82)}              && \multicolumn{2}{c}{80.419(67)}              \\
$T$ (K)                                  & 6993(200)         & 6166(157)               && 6993(200)         & 6146(152)               \\
$\Omega$                                 & 9.448(91)         & 5.474(62)               && 9.428(72)         & 5.448(45)               \\
$\Omega_{\rm in}$                        & \multicolumn{2}{c}{2.573}                   && \multicolumn{2}{c}{2.573}                   \\
$A$                                      & 1.0               & 0.5                     && 1.0               & 0.5                     \\
$g$                                      & 1.0               & 0.32                    && 1.0               & 0.32                    \\
$X$, $Y$                                 & 0.639, 0.255      & 0.639, 0.230            && 0.639, 0.255      & 0.639, 0.230            \\
$x$, $y$                                 & 0.615, 0.296      & 0.657, 0.277            && 0.615, 0.296      & 0.658, 0.277            \\
$L$/($L_{1}$+$L_{2}$)                    & 0.7276(38)        & 0.2724                  && 0.7288(15)        & 0.2712                  \\
$r$ (pole)                               & 0.1099(12)        & 0.0842(23)              && 0.1101(9)         & 0.0846(9)               \\
$r$ (point)                              & 0.1101(12)        & 0.0845(23)              && 0.1103(9)         & 0.0849(9)               \\
$r$ (side)                               & 0.1100(12)        & 0.0843(23)              && 0.1102(9)         & 0.0847(9)               \\
$r$ (back)                               & 0.1101(12)        & 0.0845(23)              && 0.1103(9)         & 0.0849(9)               \\
$r$ (volume)$\rm ^c$                     & 0.1100(12)        & 0.0843(23)              && 0.1102(9)         & 0.0848(9)               \\ 
$\sum W(O-C)^2$                          & \multicolumn{2}{c}{0.0021}                  && \multicolumn{2}{c}{0.0011}                  \\ [1.0mm]
\multicolumn{6}{l}{Absolute parameters:}                                                                                              \\            
$M$ ($M_\odot$)                          & 1.50(8)           & 0.52(3)                 && 1.50(8)           & 0.52(3)                 \\
$R$ ($R_\odot$)                          & 2.26(6)           & 1.73(6)                 && 2.26(5)           & 1.74(4)                 \\
$\log$ $g$ (cgs)                         & 3.91(3)           & 3.68(4)                 && 3.90(3)           & 3.67(3)                 \\
$\rho$ (g cm$^3)$                        & 0.18(2)           & 0.14(2)                 && 0.18(2)           & 0.14(1)                 \\
$L$ ($L_\odot$)                          & 11(1)             & 3.9(5)                  && 11(1)             & 3.9(4)                  \\
$M_{\rm bol}$ (mag)                      & 2.13(14)          & 3.26(13)                && 2.13(13)          & 3.26(12)                \\
BC (mag)                                 & 0.03(1)           & $-$0.02(2)              && 0.03(1)           & $-$0.03(2)              \\
$M_{\rm V}$ (mag)                        & 2.10(14)          & 3.28(14)                && 2.10(13)          & 3.29(12)                \\
Distance (pc)                            & \multicolumn{2}{c}{482$\pm$31}              && \multicolumn{2}{c}{482$\pm$30}              \\
\enddata
\tablenotetext{a}{Result from the observed data.}
\tablenotetext{b}{Result from the pulsation-subtracted data.}
\tablenotetext{c}{Mean volume radius.}
\end{deluxetable}

\begin{deluxetable}{lrccrc}
\tablewidth{0pt}
\tablecaption{Multiple Frequency Analysis of TIC 309658221$\rm ^a$ }
\tablehead{
             & \colhead{Frequency}    & \colhead{Amplitude} & \colhead{Phase} & \colhead{S/N$\rm ^b$} & \colhead{Remark}        \\
             & \colhead{(day$^{-1}$)} & \colhead{(mmag)}    & \colhead{(rad)} &               &
}
\startdata                                                                                             
$f_{1}$      &  9.94597$\pm$0.00002   & 2.07$\pm$0.02       & 3.99$\pm$0.02   & 208.42        &                                 \\
$f_{2}$      & 10.57708$\pm$0.00003   & 1.31$\pm$0.02       & 0.98$\pm$0.04   & 128.87        &                                 \\
$f_{3}$      & 10.80865$\pm$0.00007   & 0.59$\pm$0.02       & 5.32$\pm$0.09   &  57.69        &                                 \\
$f_{4}$      & 13.00705$\pm$0.00008   & 0.53$\pm$0.02       & 2.71$\pm$0.09   &  52.82        &                                 \\
$f_{5}$      & 12.34430$\pm$0.00010   & 0.45$\pm$0.02       & 1.42$\pm$0.11   &  44.24        &                                 \\
$f_{6}$      & 10.50692$\pm$0.00010   & 0.44$\pm$0.02       & 0.91$\pm$0.11   &  43.93        &                                 \\
$f_{7}$      &  9.91961$\pm$0.00010   & 0.41$\pm$0.02       & 3.19$\pm$0.12   &  41.29        & $f_2+f_5-f_4$                   \\
$f_{8}$      &  0.13278$\pm$0.00035   & 0.38$\pm$0.05       & 3.41$\pm$0.42   &  11.87        & $f_{\rm orb}$                   \\
$f_{9}$      & 11.44596$\pm$0.00012   & 0.36$\pm$0.02       & 1.87$\pm$0.14   &  35.50        & $f_2+f_3-f_1$                   \\
$f_{10}$     & 14.14393$\pm$0.00012   & 0.32$\pm$0.02       & 0.92$\pm$0.15   &  33.88        & $f_4+2f_6-2f_1$                 \\
$f_{11}$     &  0.41322$\pm$0.00045   & 0.30$\pm$0.05       & 4.64$\pm$0.54   &   9.26        & 3$f_{\rm orb}$                  \\
$f_{12}$     &  8.66007$\pm$0.00018   & 0.24$\pm$0.02       & 5.70$\pm$0.22   &  23.18        & 2$f_6-f_5$                      \\
$f_{13}$     & 11.95355$\pm$0.00022   & 0.19$\pm$0.02       & 1.20$\pm$0.27   &  18.77        & $f_{10}+f_3-f_4$                \\
$f_{14}$     & 16.09894$\pm$0.00020   & 0.18$\pm$0.01       & 3.77$\pm$0.24   &  21.10        & 2$f_4-f_7$                      \\
$f_{15}$     &  8.05874$\pm$0.00025   & 0.18$\pm$0.02       & 5.70$\pm$0.29   &  17.02        & $f_{12}+f_7-f_6$                \\
$f_{16}$     &  9.73286$\pm$0.00025   & 0.16$\pm$0.02       & 5.17$\pm$0.30   &  16.55        & $f_1+f_2-f_3$                   \\
$f_{17}$     &  9.14022$\pm$0.00035   & 0.12$\pm$0.02       & 1.71$\pm$0.42   &  12.02        & $f_2+f_6-f_{13}$                \\
$f_{18}$     & 12.79873$\pm$0.00040   & 0.11$\pm$0.02       & 4.09$\pm$0.48   &  10.51        & $f_{13}+f_3-f_1$                \\
$f_{19}$     &  9.65329$\pm$0.00053   & 0.08$\pm$0.02       & 2.54$\pm$0.63   &   7.91        & $f_1+f_6-f_3$                   \\
$f_{20}$     &  9.23767$\pm$0.00062   & 0.07$\pm$0.02       & 3.89$\pm$0.74   &   6.80        & $f_{19}-f_{11}$                 \\
$f_{21}$     &  9.93635$\pm$0.00375   & 0.01$\pm$0.02       & 0.67$\pm$4.46   &   1.12        & $f_1$                           \\
$f_{22}$     & 20.52151$\pm$0.00059   & 0.06$\pm$0.01       & 5.10$\pm$0.70   &   7.14        & $f_1+f_2$                       \\
$f_{23}$     & 10.30834$\pm$0.00070   & 0.06$\pm$0.02       & 1.99$\pm$0.83   &   6.03        & $f_2-2f_{\rm orb}$              \\
$f_{24}$     & 11.04152$\pm$0.00072   & 0.06$\pm$0.02       & 4.55$\pm$0.86   &   5.86        & $f_9-f_{11}$                    \\
$f_{25}$     & 19.23727$\pm$0.00065   & 0.05$\pm$0.01       & 6.17$\pm$0.77   &   6.50        & $f_{12}+f_2$                    \\
$f_{26}$     & 13.72771$\pm$0.00095   & 0.04$\pm$0.02       & 3.93$\pm$1.13   &   4.42        & $f_{10}-f_{11}$                 \\
\enddata                                                                                                                           
\tablenotetext{a}{Frequencies are listed in order of detection. }
\tablenotetext{b}{Calculated in a range of 5 day$^{-1}$ around each frequency. }
\end{deluxetable}

\end{document}